\documentclass[12pt]{article}

\usepackage{amssymb}
\usepackage{amsmath}
\usepackage{latexsym}

\catcode`@=11
\renewcommand{\section}{\@startsection{section}{1}{0pt}{\medskipamount}
{\medskipamount}{\large\bf}}
\numberwithin{equation}{section}
\catcode`@=12
\def\beq{\begin{eqnarray}}    %%%  begequation/eqnarray
\def\eeq{\end{eqnarray}}      %%%  endequation/eqnarray

\def\ln{\,\mbox{ln}\,}                  %%% log
\def\sdet{\,\mbox{sdet}\,}              %%% superdeterminant
\def\pa{\partial}                       %%% partial
\def\={\ =\ }

\begin{document}

\begin{center}

{\Large\bf Finite BRST-BFV transformations for dynamical systems
with second-class constraints}

\vspace{18mm}

{\Large Igor A. Batalin$^{(a, b)}\footnote{E-mail:
batalin@lpi.ru}$\;,
Peter M. Lavrov$^{(b, c)}\footnote{E-mail:
lavrov@tspu.edu.ru}$\;,
Igor V. Tyutin$^{(a, b)}\footnote{E-mail: tyutin@lpi.ru}$
}

\vspace{8mm}

\noindent ${{}^{(a)}}$
{\em P.N. Lebedev Physical Institute,\\
Leninsky Prospect \ 53, 119 991 Moscow, Russia}

\noindent  ${{}^{(b)}}
${\em
Tomsk State Pedagogical University,\\
Kievskaya St.\ 60, 634061 Tomsk, Russia}

\noindent  ${{}^{(c)}}
${\em
National Research Tomsk State  University,\\
Lenin Av.\ 36, 634050 Tomsk, Russia}

\vspace{20mm}

\begin{abstract}
\noindent
We study finite field dependent BRST-BFV transformations for dynamical systems
with first- and second-class constraints within the generalized Hamiltonian formalism.
We find explicitly their Jacobians and the form of a solution to the compensation
equation necessary for
generating an arbitrary finite change of gauge-fixing functionals
in the path integral.
\end{abstract}

\end{center}

\vfill

\\

\\

%\newpage
\section{Introduction}

As far as the Hamiltonian constrained dynamics is concerned, it is
well-known that one can always convert original second-class
constraints into first-class ones by introducing extra degrees of
freedom \cite{FSh,BF1,BF2,EM,BFF1,BFF3,BT1}. Thus, in
principle, one is always allowed  to deal with first-class
constraints only. However, because of some specific reasons one can
do prefer to work directly with original second-class constraints,
as they defined by Dirac \cite{Dirac,BG} (see also
models \cite{LSh,GR}). Here we recall some elementary facts as to
the construction of the path integral for the partition function in
that case. A new feature in our analysis is that the invariance of
the formalism under rotation of second-class constraints is also
shown to be a kind of a BRST symmetry in miniature.
\\

\section{Pure second-class constraints}

Let
\beq
\label{2.1}
Z^{A} =  ( P_{i}, Q^{i} ),  \quad
\varepsilon ( P_{i} ) =\varepsilon ( Q^{i} ),    %  (1)
\eeq
be a set of original canonical variables.  Let
\beq
\label{2.2}
H( Z ), \quad \varepsilon( H ) = 0,     %    (2)
\eeq
be an original non-degenerate  Hamiltonian, and let
\beq
\label{2.3}
\Theta^{\alpha}( Z ), \quad \varepsilon ( \Theta^{\alpha} ) =
\varepsilon_{\alpha},   %   (3)
\eeq
be original second-class constraints, so that their Poisson bracket
matrix,
\beq
\label{2.4}
\{ \Theta^{\alpha}, \Theta^{\beta} \},       %  (4)
\eeq
is invertible.  Let us define the action
\beq
\label{2.5}
W   =  \int dt \left[ \frac{1}{2} Z^{A} \omega_{AB} \frac{d Z^{B}}{dt} - H -
\Theta^{\alpha} \xi_{\alpha} -\frac{1}{2}
 C_{\alpha} \{ \Theta^{\alpha}, \Theta^{\beta}\} C_{\beta} \right],  %(5)
\eeq
where $\omega _{AB}$ is an inverse to
\beq
\label{2.6}
\omega^{AB} = \{ Z^{A}, Z^{B} \} = {\rm const}( Z ),         %     (6)
\eeq
$\xi_{\alpha}$ are Lagrange multipliers, $\varepsilon( \xi_{\alpha} ) =
\varepsilon_{\alpha}$, $C_{\alpha}$ are Dirac ghosts, $\varepsilon( C_{\alpha} ) =
\varepsilon_{\alpha} + 1$.

The partition function is given by the path integral
\beq
\label{2.7}
\mathcal{Z}  =  \int [DZ] [D\xi] [DC] \exp\left\{\frac{ i}{\hbar } W \right\}.  %(7)
\eeq
The action (\ref{2.5}) is invariant under the following  "BRST
transformations" with $\mu$ being a  fermionic parameter,
\beq
\label{2.8}
\delta Z^{A} = \{ Z^{A}, \Theta^{\alpha} \} C_{\alpha} \mu,      %  (8)
\eeq
\beq
\label{2.9}
\delta C_{\beta}  =  \mu \{ H, \Theta^{\alpha} \} D_{\alpha \beta}  +
\xi_{\beta} \mu,  \quad \delta \xi_{\alpha}=0,  %  (9)
\eeq
where $D_{\alpha \beta}$ is an inverse to $ \{ \Theta^{\alpha}, \Theta^{\beta}\}$.
For constant $\mu$,  the  Jacobian of the transformations  (\ref{2.8}) and (\ref{2.9})
equals to one. Thus, the path integral (\ref{2.7}) is stable under the transformations
(\ref{2.8}) and (\ref{2.9}) with
constant $\mu$.

Now, let us consider a field-dependent Fermionic parameter of the form
\beq
\label{2.10}
\mu  =\frac{i}{\hbar}  \int dt C_{\alpha} \delta
\Lambda^{\alpha}_{\beta} \Theta^{\beta},  %     (10)
\eeq
where arbitrary infinitesimal matrix $\delta \Lambda$ is $Z$-dependent.
In that case, the transformations (\ref{2.8}) and (\ref{2.9}) yield the Jacobian,
\footnote{In accordance
with the general ideology of Ref. \cite{blt1}, the transformations
(\ref{2.8}) and (\ref{2.9}) can also be generalized to the case of finite field-dependent
fermionic parameter $\mu$, although their Jacobians in that case are
modified essentially with the terms containing explicitly the differential squared,
corresponding to (\ref{2.8}), (\ref{2.9}), as
applied to the Fermionic BRST parameter $\mu$,
$\ln J=-\ln ( 1 + \kappa )-( \mu \overleftarrow{d }^2 )( 1+\kappa)^{-1} \mu,
\kappa = ( \mu \overleftarrow{d })$ (see also \cite{elr}), where the differential is
defined via the transformations (\ref{2.8}) and (\ref{2.9}) as $\overleftarrow{d} = \int
dt[(\overleftarrow{\delta}/\delta Z^{A})\delta Z^{A} +
(\overleftarrow{\delta}/\delta C_{\beta})\delta C_{\beta}](\overleftarrow{\pa}/\pa\mu)$.
However, as the latter
differential is not nilpotent, one cannot guarantee the existence  of a
solution for a finite fermionic parameter $\mu$ generating  an arbitrary  finite
rotation of the constraints (\ref{2.3}).}
\beq
\label{2.11}
J =  1  - \frac{i}{\hbar} \int dt [ \delta \Lambda^{\alpha}_{\beta}
\Theta^{\beta} \xi_{\alpha}  +
C_{\alpha} \{ \delta \Lambda^{\alpha}_{\beta} \Theta^{\beta},
\Theta^{\gamma} \} C_{\gamma}+\delta H],    %    (11)
\eeq
which induces arbitrary infinitesimal rotation of the constraints,
\beq
\label{2.12}
\Theta \;\rightarrow \; ( 1 + \delta \Lambda ) \Theta,       %   (12)
\eeq
in the integrand in (\ref{2.7}), accompanied  by  the weakly vanishing
variation of the Hamiltonian,
\beq
\label{i2.13}
\delta H = - \{ H, \Theta^{\alpha} \} D_{ \alpha \beta } \!\;\delta \Lambda
^{\beta}_{\gamma} \Theta^{\gamma}  \simeq 0.    %  (2.13) ".
\eeq
Equivalently, one can say that the $\delta H$, (\ref{i2.13}), can be compensated by
the corresponding shift of $\xi_{\gamma}$ to the first order in $\delta \Lambda$,
\beq
\label{i2.14}
\delta \xi_{\gamma}   =  -  \{ H, \Theta^{\alpha} \} D_{ \alpha \beta }\!\; \delta
\Lambda^{\beta}_{\gamma} (-1)^{ \varepsilon_{\gamma} }.       %    (2.14)
\eeq
Thus, we have confirmed that the path
integral (\ref{2.7}) is, in fact, independent of the special choice of the basis of constraints.
On the other hand, by rotating the basis, one can
always make the Poisson bracket matrix of the constraints be
constant, so that the Dirac ghosts decouple,  and the path integral reduces
explicitly to the physical degrees of freedom.

Finally, let us rewrite  the path integral in its "conceptual" form,
\beq
\label{2.13}
\mathcal{Z} = \int \exp \left\{\frac{i}{\hbar} \int dt \left[\frac{1}{2} Z^{A}
\omega_{AB} \frac{d Z^{B}}{dt }  -  H( Z ) \right] \right\} d \mu[Z],      % (12)
\eeq
with the functional measure
\beq
\label{2.14}
d \mu[Z] =  \delta [\Theta] \rho [Z] [ DZ ],   %      (13)
\eeq
where
\beq
\delta [\Theta] = \int [D \xi] \exp\left\{ -
\frac{i}{\hbar} \int dt \Theta \xi  \right\}= \prod_{t} \delta( \Theta ),
\eeq
is a functional $\delta$ - function of the constraints,
\beq
\nonumber
\rho [Z] &=& \int [DC] \exp\left\{ - \frac{i}{2\hbar}
\int dt C \{ \Theta, \Theta\}C \right\}=\\
\label{2.15}
&=&\exp\left\{ \delta ( 0 )\frac{1}{2}
\int dt  \ln\big( \sdet \{ \Theta, \Theta\} \big)\right\}=
\prod_{t} \sqrt{ \sdet \{ \Theta, \Theta  \} }%^{1/2},  %(14)
\eeq
is a measure density as represented in terms of the Pfaffian \cite{F}
(see also \cite{PS,FV2}).
Notice an important
invariance property of the measure (\ref{2.14}) under the transformations $\delta
Z^{A} = \{ Z ^{A}, G \}_{\cal D}$,
generated canonically by the Dirac bracket on the hypersurface
$\Theta^{\alpha} = 0$,
\beq
\label{2.17}
\big(\rho(Z)\big)^{-1}\pa_{A}\big(\rho(Z)\omega^{AB}_{{\cal D}}(Z)\big)
(-1)^{\varepsilon_{A}} = 0,  % (2.17)
\eeq
\beq
\label{2.18}
\omega^{AB}_{{\cal D}}(Z) = \{Z^{A}, Z^{B}\}_{{\cal D}},\quad \rho(Z) =
\sqrt{ \sdet \{\Theta, \Theta \} }%^{1/2}.  %(2.18)
\eeq
Recall that for any functions $F, G$, the Dirac bracket is defined in terms of the
Poisson brackets as
\beq
\{F, G\}_{{\cal D}} = \{F, G\} - \{F, \Theta^{\alpha}\}
 D_{\alpha\beta} \{\Theta^{\beta}, G \},
%(2.19)
\eeq
where $D_{\alpha\beta}$ is an inverse to (\ref{2.4}). The Dirac bracket satisfies the
antisymmetry, Leibnitz rule and Jacobi identity, in the same sense as usual Poisson
brackets do. The Dirac brackets also satisfy
\beq
\{F, \Theta^{\alpha}\}_{{\cal D}} = 0,   %  (2.20)
\eeq
for any $F$.

Thus, our final statement here is that there exists a similarity
between the arbitrariness in
rotation of constraints, and genuine gauge invariance.
\\

\section{First- and second-class constraints together}

For the sake of technical simplicity, in the present section, we consider
only irreducible gauge theories,  whose first-class
constraints are linearly independent, by definition.  We begin with
describing the general structure of the extended phase
space intended specifically to quantize irreducible gauge theories.  Here we
denote  by $Z^{A}$ the total set of canonical pairs
of the extended phase space,
\beq
\label{3.1}
Z^{A} = ( z^{i};  \pi_{a}, \lambda^{a};  \bar{ \mathcal{P} }_{a}, C^{a};
\bar{C}_{a}, \mathcal{P}^{a} ),  %     (3.1)
\eeq

(i) ($z^{i}$)  denotes a set of original canonical variables, their
Grassmann parities are ($ \varepsilon_{i}$ ),  their ghost numbers are ($0 $);
original Hamiltonian $H_{0}( z )$, first-class constraints $T_{a}( z )$, and
second-class constraints $\Theta_{\alpha}( z )$ are regular functions
of $z^{i}$ only, their  Grassmann parities are ($0, \varepsilon_{a},
\varepsilon_{\alpha}$ );  all other canonical  variables  are split
explicitly into pairs of canonical  momenta  and coordinates;  among the latter
canonical pairs are:

(ii)  Dynamically  active  Lagrange  multipliers  to  first-class
constraints and  to  their  gauges,
\beq
\label{3.2}
(\pi_{A}, \lambda^{a}),  % (3.2)
\eeq
their Grassmann parities are ($\varepsilon_{A},
\varepsilon_{a} $), their ghost numbers are ($0, 0)$ ;     % (3.2)

(iii)  Ghosts,
\beq
\label{3.3}
( \bar{ \mathcal{P}}_{a} , C^{a}  ), % (3.3)
\eeq
their Grassmann parities are (
$\varepsilon_{a} + 1, \varepsilon_{a} + 1 $),
their ghost numbers are ($ - 1, +1 $);    % (3.3)

(iv) Antighosts,
\beq
\label{3.4}
( \bar{C}_{a}, \mathcal{P}^{a} ),
\eeq
their Grassmann parities are  ($\varepsilon_{a} + 1,  \varepsilon_{a} + 1 $),
their ghost numbers are ($ - 1,+ 1 $).   %  (3.4)

We proceed with the original  Dirac bracket form of the classical gauge
algebra,
\beq
\label{3.5}
\{ T_{a}, T_{b} \}_{{\cal D}}\simeq  U_{ab}^{c}T_{c},  \quad
\{ T_{a}, H_{0} \}_{{\cal D}} \simeq V_{a}^{b} T_{b},    %   (3.5)
\eeq
where $\simeq$ means weak equality, modulo arbitrary linear combination of
second-class constraints $\Theta^{\alpha}$ \cite{Dirac}.

Given the classical gauge algebra (\ref{3.5}),  one defines the fermionic BRST-BFV
generator $\Omega$ and bosonic extended
Hamiltonian $\mathcal{H}$,  to satisfy the  gauge-algebra generating
equations,
\beq
\label{3.6}
\{ \Omega, \Omega \}_{{\cal D}} \simeq 0 ,  \quad    \varepsilon( \Omega ) =  1,
\quad {\rm gh}( \Omega ) = 1,     %  (3.6)
\eeq
\beq
\label{3.7}
\{ \Omega,  \mathcal{H} \}_{{\cal D}} \simeq 0,\quad \varepsilon( \mathcal{H} ) = 0,
\quad {\rm gh}( \mathcal{H} ) = 0.  %    (3.7)
\eeq
The existence of a solution \cite{BT1,FF,Tyutin,BT2} to  the gauge algebra  generating  Eqs.
(\ref{3.6}) and  (\ref{3.7})  is guaranteed  by  the following
consequences of the Jacobi identities  for  the Dirac  brackets,
\beq
\{ \{ F,  F \}_{\cal D},  F \}_{\cal D}  =  0,  \quad
 \{ \{X,  F\}_{\cal D},  F \}_{\cal D}   =  \{ X,  (1/2)
\{ F,  F\}_{\cal D} \}_{\cal D} , \quad
\varepsilon(F)  =  1, \quad  {\rm any}\;\;  X .    %  (3.8)
\eeq
One  has to  seek for a solution to these generating  equations in the form
of a ghost power series expansions,
\beq
\label{3.8}
\Omega  \simeq  \mathcal {P}^{a} \pi_{a} + \left[  C^{a} T_{a} +
\frac{1}{2} (-1)^{ \varepsilon_{ b } } C^{b} C^{a} U_{ab}^{c} \bar{
\mathcal{P} }_{c} (-1)^{ \varepsilon_{c} } +
\mathcal{O}( CCC \bar{\mathcal{P} } \bar{ \mathcal{P} } )  \right],   %(3.8)
\eeq
\beq
\label{3.9}
\mathcal{H}\simeq H_{0} + C ^{a} V_{a}^{b} \bar{ \mathcal{P}}_{b} (-1)^{
\varepsilon_{b} } +
\mathcal{O} ( CC \bar{\mathcal{P}} \bar{ \mathcal{P} } ).   %   (3.9)
\eeq
Respectively, to the $CC$ - and $C$- order,  Eqs. (\ref{3.6})
and (\ref{3.7}) reproduce the gauge algebra relations (\ref{3.5}).
Higher structure relations of the gauge algebra are reproduced to higher
orders in ghosts.

Define the complete unitarizing  Hamiltonian $H_{\Psi}$  by the formula
\beq
\label{3.10}
H_{\Psi} \simeq \mathcal{H}  + \{ \Omega , \Psi \}_{\cal D}, \quad
\varepsilon(\Psi) =1, \quad  {\rm gh}( \Psi ) =  - 1 .    %   (3.10)
\eeq
where $\Psi$ is a gauge-fixing Fermion function of the form
\beq
\label{3.11}
\Psi  \simeq  \lambda^{a} \bar{ \mathcal{P} }_{a}  +  \chi^{a}
\bar{C}_{a},     %     (3.11)
\eeq
with $\chi^{a}$ being just the gauge functions by themselves. They are allowed
to depend on all the phase variables, under the only condition that
\beq
\label{3.12}
{\rm gh}( \chi^{a} ) = 0.    %      (3.12)
\eeq

Due to (\ref{3.6}) and (\ref{3.7}),
\beq
\label{3.13}
\{ H_{\Psi}, \Omega \}_{\cal D} \simeq 0.    %  (3.13)
\eeq
To  the  second  order in ghosts,  with  $\chi^{a} = \chi^a(z;\pi,\lambda)$,  the
unitarizing  Hamiltonian is
\beq
\nonumber
H_{\Psi} &=& H_{0} + \big( T_{a}  + (-1)^{ \varepsilon_{a} } C^{b} U_{ba}^{c}
\bar{ \mathcal{ P } }_{c} (-1)^{ \varepsilon_{c} } \big) \lambda^{a} +
\pi_{a} \chi^{a}  + \\
\label{3.14}
&&+ \bar{C}_{a} \{ \chi^{a}, T_{b} \}_{\cal D} C^{b}  +
\bar{C}_{a} \{ \chi^{a}, \pi_{b} \}_{\cal D} \mathcal{ P }^{b} +
\big( C^{a}V_{a}^{b}  - \mathcal{P}^{b}\big) \bar{\mathcal{ P } }_{b} (-1)^{
\varepsilon_{b}} .         %         (3.14)
\eeq
Now, define the complete action,
\beq
\label{3.15}
W_{\Psi}  =  \int dt \left[\frac{1}{2} Z^{A}\omega_{AB} \frac{d Z^{B}}{dt}
-  H_{\Psi} \right ], %(3.15)
\eeq
in terms of the unitarizing  Hamiltonian  (\ref{3.10}).  Then, we define the
corresponding path integral \cite{FF},
\beq
\label{3.16}
\mathcal{Z} = \mathcal{Z}_{\Psi}  =
\int  \exp\left\{\frac{i}{\hbar}  W_{\Psi} \right\} d\mu[Z],  %    (3.16)
\eeq
with the functional measure
\beq
\label{3.17}
d\mu[Z] = \delta[\Theta] \rho[z] [DZ],  %   (3.17)
\eeq
where
\beq
\label{3.18}
\delta[\Theta] = \int [D\xi] \exp\left\{ -\frac{i}{\hbar} \int dt  \Theta^{ \alpha } (
z ) \xi_{\alpha}\right \} =
\prod_{t}\delta( \Theta ( z )) ,   %   (3.18)
\eeq
\beq
\nonumber
&&\rho[z] = \int [DC] \exp\left\{ -\frac{i}{\hbar }
\int dt \frac{1}{2} C^{\alpha} \{
\Theta^{\alpha}( z ), \Theta^{\beta}( z ) \} C^{\beta} \right\} =\\
\label{3.19}
&&
= \exp\left\{ \delta( 0 ) \int dt \ln \rho( z ) \right \}  = \prod_{t} \rho( z )
,  \quad \rho(z) =  \sqrt{ \sdet \{ \Theta( z ), \Theta( z )\} }. %(3.19)
\eeq
In analogy with (\ref{2.17}), one has in the sector of the original variables
$z^{i}$, \cite{FF},
\beq
\label{3.20}
( \rho( z ) )^{-1} \pa_{i} ( \rho( z ) \omega_{\cal D}^{ij}( z ) )
(-1)^{\varepsilon_{i}} = 0.  %   (3.20)
\eeq

In the path integral  (\ref{3.16}), consider the infinitesimal BRST-BFV
transformation,
\beq
\label{3.21}
\delta Z^{A} \simeq \{ Z^{A}, \Omega \}_{\cal D} \mu.    %   (3.21)
\eeq
On the constraint surface
\beq
\label{3.22}
\Theta^{\alpha}( z ) = 0, \quad ( {\rm any}\;\; t ) , %   (3.22)
\eeq
the induced variation in the action (\ref{3.15}) is given by the boundary term,
\beq
\label{3.23}
\delta W_{\Psi} = \left[\frac{1}{2}\big( Z^{A} P_{A}^{B}\pa_{B}-2 \big)
\Omega\mu\right] \Big|_{-\infty}^{ +\infty },   % (3.23)
\eeq
where we have denoted the Dirac projector matrix
\beq
\label{3.24}
P_{A}^{B} = \omega_{AC}\!\; \omega_{\cal D}^{CB} .  %  (3.24)
\eeq
As to the Jacobian of the transformation (\ref{3.21}), we have
\beq
\label{3.25}
J_{\cal D} = 1 + \int dt \left[ - \mu \overleftarrow{ d_{\cal D} } +
\delta( 0 ) (-1)^{\varepsilon_{A}}( \pa_{A} \omega_{\cal D}^{AB} )
(\pa_{B}\Omega)\mu  \right],   %  (3.26)
\eeq
where
\beq
\label{3.26}
\overleftarrow{ d_{\cal D}} = \int dt \overleftarrow{\frac{\delta }{\delta Z^{A}}}
\{Z^{A}, \Omega\}_{\cal D},     %   (3.27)
\eeq
is the Dirac version of the BRST-BFV differential,
\beq
\label{3.27}
( \overleftarrow{d_{\cal D}} )^{2} \simeq 0, \quad
\varepsilon(\overleftarrow{d_{\cal D}}) = 1,\quad
{\rm gh}( \overleftarrow{d_{\cal D}}) = 1.  %   (3.28)
\eeq
Here and below in operator-valued  weak equalities,  we mean "normal -
ordered"  linear combinations of  second-class constraints,
with  functional derivative operators applying
to the left,  standing  to the left of all  the rest factors, in every
monomial. Due to the relations (\ref{2.17}) or (\ref{3.20}),
the second term in the square bracket
in the right-hand side in (\ref{3.25}),
is compensated exactly by the induced BRST-BFV variation in the density
$\rho$.

It follows then from (\ref{3.23}) and (\ref{3.25}) that the path integral (\ref{3.16})
is stable under the transformations  (\ref{3.21})
with $\mu = {\rm const}$, in case of appropriate boundary condition imposed for
integration trajectories. On the other hand, if one chooses $\mu$ in the form
\beq
\label{3.28}
\mu = \frac{i}{\hbar } \int dt \delta\Psi,   %   (3.29)
\eeq
then the Jacobian (\ref{3.25}) yields effective change of the gauge fermion,
\beq
\label{3.29}
\Psi \;\rightarrow\; \Psi + \delta \Psi.       %  (3.30)
\eeq
Thereby one has confirmed  the formal gauge independence as to the path
integral (\ref{3.16}).
\\

\section{Finite BRST-BFV transformations, their Jacobians\\ and compensation
equation}

Here, we proceed with the finite  BRST- BFV transformations in their
Dirac-bracket version,
\beq
\label{4.1}
\bar{Z}^{A} \simeq  Z^{A} + \{ Z^{A}, \Omega \}_{\cal D} \mu  =
Z^{A}( 1 + \overleftarrow{d_{\cal D}} \mu ),   %  (4.1)
\eeq
\beq
\label{i4.2}
[ \overleftarrow{d_{\cal D}} \mu_{1}, \overleftarrow{d_{\cal D}} \mu_{2} ] \simeq
\overleftarrow{d_{\cal D}} \mu_{ [12] } ,\quad
\mu_{ [12] } \simeq  - ( \mu_{1}\mu_{2} )\overleftarrow{d_{\cal D}}.   %   (4.2)
\eeq
By exactly the same reasoning as in Ref. \cite{blt1},  see  (2.19) and
(2.20) therein,  it follows that the Jacobian of the transformation (\ref{4.1})
has the general form
\beq
\label{4.2}
\ln J_{\cal D}  \simeq  - \ln( 1 + \kappa_{\cal D} ) +
\delta(0)  \int dt  (-1)^{\varepsilon_{A} }( \pa_{A} \omega_{\cal D}^{AB} )
(\pa_{B}\Omega) \mu,   %     (4.2)
\eeq
where
\beq
\label{4.3}
\kappa _{\cal D} =  \mu  \int dt \overleftarrow{\frac{\delta}{ \delta Z^{A} }}
\{Z^{A},\Omega \}_{\cal D} = \mu \overleftarrow{d_{\cal D}} .   %    (4.3)
\eeq
To the first other in $\mu$, ( i.e. in the infinitesimal case ) (\ref{4.2} ) does
coincide with (\ref{3.25}).  The same as in the latter  case,
the  second term in the right-hand side in (\ref{4.2}) is compensated, due to
(\ref{2.17}) or (\ref{3.20}),  by the BRST-BFV variation of the
density  $\rho$ in the functional measure $d\mu[Z]$. Thus, the first term in the
right-hand side in (\ref{4.2}), is the only contribution
to formulate the compensation equation as to the path integral  (\ref{3.16}),
\beq
\label{4.5}
\mu \overleftarrow{d_{\cal D}}\simeq   \exp\left\{\frac{i}{ \hbar }
\left( \delta \Psi [Z] \overleftarrow{d_{\cal D}}\right) \right\}  -  1,\quad
\Psi [Z]  = \int dt  \Psi( Z(t) ).  %(4.5)
\eeq
An obvious solution to that equation has the form
\beq
\label{4.6}
\mu[ \delta\Psi ]= \frac{i}{\hbar} E\big( (i/\hbar) ( \delta\Psi [Z]
\overleftarrow{d_{\cal D}}) \big)\delta\Psi[Z],
\quad  E( x )  =  x^{-1} ( \exp x - 1 ).     %    ( 4.6)
\eeq
If one chooses the variables (\ref{4.1}) with parameters (\ref{4.6}), to be the new
integration variables in the path integral (\ref{3.16}),
then, in the new variables, one gets the new gauge fermion,  $\Psi_{1} =
\Psi  + \delta \Psi$.
If one introduce external source $J_{A}( t )$, to define the generating
functional,
\beq
\label{i4.6}
\mathcal{ Z }_{ \Psi } = \int d\mu [ Z ] \exp\left\{\frac{i}{\hbar}  \left[ W_{ \Psi } +
\int dt  J_{A} Z^{A} \right] \right\},     % (4.6)
\eeq
then the following interpolation formula between the two finite-differing
gauges,  $\Psi_{1}$  and  $\Psi$,  holds
\beq
\label{4.7}
\mathcal{ Z }_{\Psi_{1}}  =  \mathcal{ Z }_{ \Psi } \left[1 +
{\Big\langle} \frac{i}{\hbar } \int dt  J_{A} \left( Z^{A}\overleftarrow{d_{\cal D}}\right)
\mu[ - \delta\Psi ] {\Big\rangle}_{ \Psi }  \right] ,   %  (4.7)
\eeq
where $\mu[\delta\Psi ]$ is given by (\ref{4.6}),  and  the quantum mean value,  $\langle
( ... ) \rangle_{ \Psi }$,  is defined by
\beq
\label{4.7}
\langle( ... )\rangle_{\Psi}  =  ( \mathcal{Z}_{ \Psi } )^{-1}
\int d\mu[ Z ] ( ...) \exp\left\{\frac{i}{\hbar} \left[ W_{\Psi}  +
\int dt  J_{A} Z^{A}\right] \right\}.    %  (4.8)
\eeq
Thus, one has confirmed
that finite BRST-BFV transformations in their  Dirac-bracket version are
quite capable of inducing finite change of gauge-fixing fermion in
the path integral in the presence of second-class
constraints.  As the situation with the other aspects
of the matter is quite obvious, we have no reason to consider the aspects
here in further detail (see \cite{blt1}).
\\

\section{Discussion}
\noindent
In the present article, we have extended our study \cite{blt1} of finite field-dependent
BRST-BFV
transformations within the generalized Hamiltonian formalism \cite{FVh1,BVh}, to the case of
the second-class constraints present. It was shown that the invariance of the formalism
under rotations of second-class constraints can be represented in the form of a
BRST-like symmetry. An explicit form of the Jacobian of the finite BRST-BFV transformation
was found in terms of the Dirac-bracket version of the functional differential
applying on the space of trajectories. We have formulated the compensation equation
determining the finite parameter of the BRST-BFV transformation to make its Jacobian yield
arbitrary finite change in the gauge-fixing fermion function. It was confirmed that all the
results of \cite{blt1} generalize naturally via replacement of:  (i), the ordinary Poisson
bracket by the Dirac bracket, (ii) the trivial canonical integration measure in the path
integral by the Dirac measure, and  (iii) considering all basic equations in the weak sense
of Dirac.

In conclusion, we demonstrate how  the "conceptual" form  (\ref{2.13}) -
(\ref{2.15}) of the path integral with second-class constraints does
generalize as to the case of the general coordinates $Z^{A}$,
whereas the basic invertible  symplectic metric,
\beq
\label{5.1}
\omega^{AB} (Z )  =  \{ Z^{A}, Z^{B } \},           %    (5.1)
\eeq
is not a constant in $Z$.  The latter metric,  in its  contravariant
components, does satisfy the Jacobi identity,
\beq
\label{5.2}
\omega^{AD} \pa_{D} \omega^{BC} (-1)^{ \varepsilon_{A}
\varepsilon_{C} }  + {\rm cyclic \;\; perm.}( A,B,C )  =  0,   %    (5.2)
\eeq
or,  in its  covariant components $\omega_{AB},\;\;
\omega^{AB}\omega_{BC} = \delta^{A}_{C}$,  one has,
\beq
\label{5.3}
\pa_{C} \omega_{AB} (-1)^{ ( \varepsilon_{C} + 1 ) \varepsilon_{B} }
+ {\rm cyclic \;\; perm.}( A,B,C)  =  0.         %      (5.3)
\eeq
Then, one should make the following replacements  in (\ref{2.13}) and (\ref{2.14}).
In the integrand in  (\ref{2.15}),  in the
exponential, in the square brackets,  one should replace
\beq
\label{5.4}
\frac{1}{2} \omega_{AB} \;\rightarrow \; {\bar\omega}_{AB}  =
( Z^{C} \pa_{C} + 2)^{-1} \omega_{AB},      %  (5.4)
\eeq
in the  kinetic part of the action.  In  (\ref{2.14}),  one  should
replace
\beq
\label{5.5}
\rho [Z]\;\rightarrow  \; \bar{\rho} [Z]  =   \exp\left\{ \delta (0) \int dt
\ln (\bar{\rho } ( Z ) )\right\}  = \prod_{t} \bar{\rho } ( Z ) ,    %  (5.5)
\eeq
where the new local density is given by
\beq
\label{5.6}
{\bar \rho } ( Z ) = \rho ( Z )  \sqrt{\sdet ( \omega_{AB} )},  % (5.6)
\eeq
via  the "old" local density, the second in (\ref{2.18}).   In the
canonical invariance property (\ref{2.17}), one should replace
\beq
\label{5.7}
\rho ( Z )\;\rightarrow \; \bar{ \rho } ( Z ).        %    (5.7)
\eeq

As to the general-coordinate version of the path integral  (\ref{3.16}),  with
the  first-class constraints  present,
the latter generalization, in principle,  includes  the  same  two  steps:
one should modify the kinetic part
of the action  (\ref{3.15})  via  (\ref{5.4}),  and  the  integration  measure (\ref{3.17}) -
(\ref{3.19})  via   (\ref{5.5}) and  (\ref{5.6}).

It should be also mentioned here that the general-coordinate version of the
constrained dynamics generalizes further to the level of a superfield \cite{BBD,BBD1},
$Z^{A}( t, \tau )  =  Z^{A}_{0}( t ) +\tau Z^{A}_{1}( t ), \varepsilon (\tau ) = 1 $,
with the covariant derivative $ D = ( d / d\tau) + \tau ( d /d t )$,  $D^{2} =
( d / d t )$,  and original action of the form
$W = \int d t d\tau [ Z^{A} \bar{ \omega }_{AB} D Z^{B}
(-1)^{\varepsilon_{B} }  -  Q _{\Psi}  ]$ and $[ Q_{\Psi}, Q_{\Psi} ]
= 0$.  Although we do not go into detail  here, note that finite BRST- BFV
transformations do correspond, within the superfield formalism,
to finite supertranslations along the $\tau$
direction. In  the superfield  path  integral,   the  superfield  delta-functional  of
second-class constraints  is included  with  trivial  ( constant ) measure density;
nontrivial density in the original phase space
is generated automatically when  getting  back to a component formalism.
\\

\section*{Acknowledgments}
\noindent
I. A. Batalin would like  to thank Klaus Bering of Masaryk
University for interesting discussions. The work of I. A. Batalin is
supported in part by the RFBR grants 14-01-00489 and 14-02-01171.
The work of P. M. Lavrov is supported by the Ministry of Education and Science of
Russian Federation, grant TSPU-122. The work of
I. V. Tyutin is partially supported by the RFBR grant 14-02-01171.
\\

%\newpage

\end{document}